# Investigation of fermionic pairing on two- dimensional tight binding lattice under phonon and electronic mechanisms within a simple Fermi liquid-like scenario and beyond- possible consequences for superconductivity in overdoped cuprates

Soumi Roy Chowdhury and Ranjan Chaudhury

Department of Condensed Matter Physics and Material Sciences

S N Bose National Centre for Basic Sciences,

Saltlake, Sector-III, Block- JD, Kolkata-700098, India

E-mail: soumi@bose.res.in, ranjan@bose.res.in

**Abstract:** Cooper's one pair problem is investigated for a 2D lattice in the background of both passive and active Fermi sea in a weakly correlated environment. Boson exchange mechanisms involving excitons as well as phonons are invoked for pairing in the s- wave channel. The important quantities calculated are pairing energy and coherence length as functions of bosonic energy, attractive coupling constant and band filling factor. Comparison of our theoretical results with those from experiments on overdoped cuprate superconductors and other type of theoretical calculations show electronic mechanism to be the more likely mechanism of pairing in the overdoped phase.

**Keywords:** A. Layered cuprates B. Superconductivity in overdoped phase C. Electronic pairing mechanism D. Fermi liquid theory E. Eliashberg theory and McMillan equation F. Passive Fermi sea and Active Fermi sea

**Introduction:**

The high critical temperature (high- $T_c$) cuprate superconductors are today quite extensively studied layered systems. They undergo transitions between different phases- from insulating with long range antiferromagnetism to superconductor with normal phase of non-Fermi liquid nature and then again to superconductor having Fermi liquid (FL) characterized normal phase, with increase in doping percentage and lowering of temperature [1]. It has been argued that with increasing concentration of doped holes the electron correlation in the system becomes weaker compared to the band width and the cuprate system becomes a better metal [2]. In contrast to the existence of isolated Fermi arcs in the underdoped phases, good and large Fermi surfaces are observed in the overdoped phases [2,3]. Again several researchers emphasized on its Fermi liquid nature following the observation of $T^2$ temperature dependence of resistivity in the overdoped phase [4]. Theoretical demonstrations of superconductivity in overdoped cuprates supported by FL description of parental normal phase too, were presented by many researchers till now [5-10]. By measuring the transport of both heat and charge in the normal state at very low temperature, experimentalists were able to verify that one of the hole doped cuprates in the over doped regime obeys the Wiedemann- Franz (WF) law, which is quite a definitive signature of FL theory [5]. The specific heat and magnetic susceptibility at the ideal composition are proportional and constant respectively with respect to temperature, consistent with Fermi liquid behavior [4]. Inspite of this simple nature of the overdoped phase, the microscopic pairing mechanism for superconductivity even in this phase is still unclear.

Motivated by the above observations and proposals, in this paper we investigate the feasibility of Cooper pairing on a realistic 2D tight binding lattice under both phonon and exciton mediated attractive interaction mechanisms separately in a FL- like background. For simplicity we consider here only s- wave pairing scheme. We then try to apply our theoretical results to the overdoped cuprate superconductors by making use of a hypothesis, to be stated later in this section. We undertake a quantitative comparison between our theoretical results obtained from these two mechanisms as well as with the results obtained from experiments. We first handle only isolated one pair problem in the background of a passive Fermi sea. Later extending our approach further, we consider the problem of pair formation in the presence of an active sea with multiple pairs. This is in spirit of making our calculations applicable to more realistic situations [11].

Toyazawa et al in 1966 obtained theoretically a criterion for the existence of a well defined and localized exciton in a square well potential circumference formed by a linear chain of ions [12]. Much later Varma adopted this model for cuprate superconductors and showed theoretically that a charge transfer (CT) exciton indeed arises between the antibonding and bonding bands formed by $Cu_{dx2-y2}$ and $Op_x p_y$ orbitals due to the constant charge fluctuation between the $Cu^{+2}O^{-2}$ and $Cu^{+1}O^{-1}$ states in the $CuO_2$ plane [13]. The presence of this exciton is manifested in the optical absorptivity graph as an additional non – Drude hump carrying a significant portion of spectral weight besides the usual Drude one [13]. It was further pointed out that CT exciton can play an important role in mediating pairing in these layered materials [13]. Some of the experimental studies also imply that the strength of pairing is determined by the coupling of the charge carriers to the unusual electronic excitations in the normal phase, suggesting that strong electron-electron interactions rather than electron- low energy boson (like phonons) interactions are more responsible for superconductivity in cuprates [14]. The possibility of the existence of electronic pairing mechanism is further strengthened following the observation that electronic inhomogeneity doesn't play a significant role in the loss of superconductivity and superfluid density in overdoped cuprates [8]. Besides, support for substantial s-wave component in the pairing in overdoped cuprates came from the investigations by various other theoretical and experimental groups using Raman scattering and tunneling spectra [15- 19].

Utilising the results from above experimental and theoretical research, our microscopic approach can be built up based on the following hypothesis:- the phenomenon of superconductivity arises in layered cuprate systems predominantly as a combination of two distinct processes viz. Cooper pair formation in individual Cu-O layers accompanied by inter layer pair tunneling.

With this calculational set up we also aim to identify the likely microscopic mechanism for superconducting pairing in the overdoped phase, by investigating the intra- layer pair formation under different conditions.

**Mathematical Formalism:**

It is almost impossible to extract a quantitative expression relating the doping concentration in the materials under consideration and the filling factor ($\delta$) of the band in our calculation. We therefore follow a scheme with only a known boundary condition that at doping concentration of 100%, the band is completely empty ($\delta=0$) i.e zero filled. Then introduction of carrier raises the degree of band filling with $\delta=1$ representing half filled band. So the lower portion of the band represents extremely overdoped region as per the phase diagram (ordering temperature vs doping concentration) of the cuprate superconductors [4]. The very upper portion of the band on the other hand corresponds to the underdoped region of cuprates. The region in between these two represents the moderated overdoped region where superconductivity occurs in a Fermi liquid like background. However drawing a boundary line separating these different regions is quite difficult.

Our calculation starts with the two-particle Hamiltonian on a 2D tight binding lattice containing the kinetic energy terms and attractive interaction ($V$) of contact type. Following the original Cooper's treatment for one pair problem in continuum in the background of a passive Fermi sea, the basic pairing equation is

$$\left\{-\left(\frac{\hbar^2}{2m}\right)(\nabla_1^2 + \nabla_2^2) + V(\vec{r}_1,\vec{r}_2)\right\}\Phi(\vec{r}_1-\vec{r}_2) = E\Phi(\vec{r}_1-\vec{r}_2) \qquad (1a)$$

where, $\Phi(\vec{r}_1 - \vec{r}_2)$ is the spin singlet pair wave function in relative co ordinate space given by, $\Phi(\vec{r}_1 - \vec{r}_2) = \sum_{\vec{k}} a_{\vec{k}}\, e^{i\vec{k}.\vec{r}_1} e^{-i\vec{k}.\vec{r}_2}$ [Corresponding to zero centre of mass momentum pairs] and E being the energy eigenvalue of the pair. The pairing energy equation for Cooper pair with zero-centre of mass momentum corresponding to tight binding lattice takes the following form in 2D [20]:

$$\left(\frac{u}{A}\right) = \left(\frac{\pi^2 B}{2}\right)\frac{1}{\int_0^{\hbar\omega_{boson}}\dfrac{K\sqrt{1-\left(\frac{\tilde{\epsilon}_k + \epsilon_F - \epsilon_0}{2t}\right)^2}\, d\tilde{\epsilon}_k}{(|W| + 2\tilde{\epsilon}_k)}} \qquad (1b)$$

where $\left(\frac{2}{\pi^2 B}\right) K\sqrt{1-\left(\frac{\tilde{\epsilon}_k + \epsilon_F - \epsilon_0}{2t}\right)^2}$ is the 2D Density of States (DOS), with $\tilde{\epsilon}_k = \epsilon_k - \epsilon_F$; $\epsilon_F$ is the Fermi energy. The quantity $V(\vec{r}_1,\vec{r}_2) = (-u\delta(\vec{r}_1 - \vec{r}_2))$ with u>0, is the attractive contact interaction whose fourier transform being equal to $(-u/A)$ only within the small energy transfer region $\hbar\omega_{boson}$ (the characteristic energy of the boson mediating attraction) above the Fermi surface (rather Fermi curve) where the pairing would take place (usual Cooper's model); A is the area of the 2D system in consideration ($A \to \infty$ for a macroscopic system). Here K denotes the complete elliptic function of the first kind; $|W|$ (with $E - 2\epsilon_F = -|W|$) is the pairing energy for the two electrons constituting a Cooper pair and B is the half band width 4t with 't' as the nearest neighbor hopping amplitude; $\epsilon_k$ being the single electron energy with the 2D lattice dispersion. It may be remarked that our lattice Hamiltonian for Cooper pairing looks somewhat similar to negative - U Hubbard model, although the attraction here operates only within a finite energy interval. It may be remarked here that to start with we keep both the possibilities of bosonic mechanism viz. electronic and phononic open under s-wave pairing scheme.

Now to examine the more realistic concept of pair formation in the presence of an active Fermi sea, we have to include the effect of Pauli's exclusion principle or rather Pauli blocking of the phase space. For this a factor $1 - f_{-k+q/2} - f_{k+q/2}$ has to be incorporated in the 2D single pairing Hamiltonian (equation 1b) where $f_{k+q/2}$ is the probability that there is a carrier of momentum $\hbar(k+\frac{q}{2})$ above the Fermi level [11]. Therefore the modified equation for pairing with finite centre of mass momentum $\hbar q$ becomes

$$1 = U \sum_{\vec{k'}} \frac{1 - f_{-k+\frac{q}{2}} - f_{k+\frac{q}{2}}}{-E + 2\epsilon_k} \qquad (1c)$$

This leads to following equation

$$1 = U \sum_{\vec{k'}} \frac{-1 + \theta\left(\epsilon_{k+\frac{q}{2}} - \epsilon_F\right) + \theta(\epsilon_{-k+\frac{q}{2}} - \epsilon_F)}{(|W| + 2\widetilde{\epsilon}_k)}$$

The characteristic of the Theta function determines the feasible range of pairing in momentum space and finally the pairing equation becomes,

$$\left(\frac{u}{A}\right) = \left(\frac{\pi^2 B}{2}\right) \frac{1}{-\int_0^{\hbar\omega_{boson}} \frac{K\sqrt{1 - \left(\frac{\widetilde{\epsilon}_k + \epsilon_F - \epsilon_0}{2t}\right)^2} d\widetilde{\epsilon}_k}{(|W| + 2\widetilde{\epsilon}_k)} + \int_{atqSin(ka)}^{\hbar\omega_{boson}} \frac{K\sqrt{1 - \left(\frac{\widetilde{\epsilon}_k + \epsilon_F - \epsilon_0}{2t}\right)^2} d\widetilde{\epsilon}_k}{(|W| + 2\widetilde{\epsilon}_k)} + \int_{-atqSin(ka)}^{\hbar\omega_{boson}} \frac{K\sqrt{1 - \left(\frac{\widetilde{\epsilon}_k + \epsilon_F - \epsilon_0}{2t}\right)^2} d\widetilde{\epsilon}_k}{(|W| + 2\widetilde{\epsilon}_k)}} \qquad (1d)$$

Generalised 2D energy dispersion can be expressed in tight binding representation as:
$\epsilon_k = \epsilon_0 - 2t(\cos(k_x a) + \cos(k_y a)) + 4t'(\cos(k_x a).\cos(k_y a)) - 2t''(\cos(2k_x a))$ (2a)
where t, $t'$ and $t''$ are the single electron hopping parameters corresponding to the nearest neighbour, next nearest neighbour and the 3$^{rd}$ neighbour respectively on a square lattice. Pavarini et al identified an intriguing correlation between $T_c$ and the ratio $t'/t$ for a large number of cuprate systems. Following this comparatively lower $T_c$ cuprates like $La_{2-x} Sr_x CuO_4$ (LSCO) and $Ba_2 Sr_{2-x} CuO_6$ (Bi-2201), with a diamond like Fermi surface at half filling for having a low value of $t'/t$ ratio, can be represented by a energy dispersion with the next near neighbour hopping ignored [21,22] :

$$\epsilon_k = \epsilon_0 - 2t(\cos(k_x a) + \cos(k_y a)) \qquad (2b)$$

It is however worthwhile to point out here that cuprates with orthorhombic structure such as YBCO and Tl based cuprates have a very large $'t'$. This aspect will be included in our calculations to get more refined results in near future.

For the numerical part of our calculations we have depended mostly on the fitted values of different parameters of cuprates such as the bosonic energy involved in the pairing process, used by Newns et al in their theoretical model, due to scanty experimental data [23, 24]. The graphs have been drawn using equations 1b and 1d to understand the behavior of |W| with variation of different parameters of cuprate superconductors corresponding to both the situations viz. passive Fermi sea and active Fermi sea. The attractive coupling constant ($\lambda$) has been calculated using the formula

$$\lambda = \left(\frac{u}{A}\right) N(\epsilon_F) \qquad (3)$$

It may be noted that although $N(\epsilon)$ is a variable here, to have an idea about the strength of the coupling, different values of u/A for a particular band filling are multiplied by the DOS at the Fermi energy corresponding to that filling [20].

Besides, estimates of coherence length for the situations corresponding to both passive and active Fermi sea, have been obtained by extending equation (1b) to finite centre of mass momentum case and equation (1d) to zero centre of mass momentum case respectively [20].

**Calculations and results:**

The whole calculation is done so that the basic Fermi- Liquid like criterion (U< 4t) is maintained. The relevant parameters presented in Table-1 are quite realistic [25].

Table-1: Parameters corresponding to pairing mechanism

| Parameters | Electronic mechanism | Phononic mechanism |
|---|---|---|
| Bosonic energy ($\hbar\omega_{boson}$) | 0.3 ev | 0.05 ev |
| Hopping amplitude ('t') | 0.2 ev | 0.2 ev |

**Single pair (passive Fermi sea):** In conventional 3D isotropic Bardeen- Cooper- Schrieffer (BCS) superconductors, an energy gap $\Delta_{sc}$ opens below Tc with s-wave symmetry and $2\Delta_{sc}$ is the minimum energy required to break a Cooper pair. The magnitude of $2\Delta_{sc}$ is of the order of 50- 60 mev for overdoped cuprates [26, 27]. Keeping this in mind, throughout our calculation the maximum value of |W| has been kept below 0.05 ev. The result of our calculations, shown in Figure 1b, indicates that the coupling constant lies in the intermediate range for electronic mechanism i.e. $\lambda \sim 0.5$. For the same range of |W| the coupling constant lies in the strong range, i.e. $\lambda \sim 0.7$ in case of phononic mechanism. Figure1a. is another graph for electronic mechanism which shows the variation of coupling constant for a higher range of |W|, where a higher coupling constant (here upto 0.65) can be obtained for this mechanism. But increasing of |W| up to this level is not possible for phononic mechanism as this violates the FL criterion.

The above results of ours are very close to those obtained by using the conventional Cooper equation corresponding to the isotropic 3D case, viz.

$$|W| = 2\hbar\omega_{boson} \exp\left(-\frac{2}{\lambda}\right) \qquad (4)$$

From the above equation (4) the value of 0.007 ev for |W| gives a coupling constant of 0.45 in case of electronic mechanism. Graphically the above magnitude of |W| corresponds to a coupling constant of 0.5 from our calculation (Figures: 1a). Furthermore, |W| of 0.007 ev corresponds to a coupling constant of 0.7 from our calculation and 0.75 from equation (4) (Figure: 1b) for phononic mechanism. We however consider the usage of this equation (4) here to be very limited as this equation is appropriate for 3D isotropic system, whereas the experimental system is truly quasi-2D or rather layered and our calculation is for pure 2D lattice. Nevertheless, we still compare the estimates of the coupling constant from the two approaches. The maximum value of the pairing energy can be upto the order of the superconducting gap according to the conventional theory of superconductivity [15, 28].

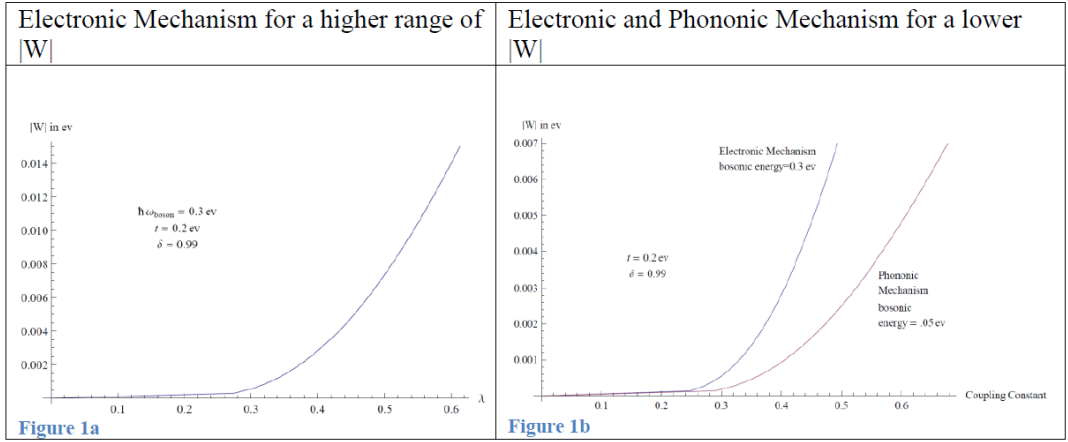

Figure1: Graphs showing variation of pairing energy with coupling constant (our theoretical results)

Some numerical results are presented below to understand the spatial nature of the pair wave function by looking at the finite centre of mass momentum case (see Table:-2). The maximum allowed pairing wave vector '$q_{max}$' (defined by $|W|_{q=0}$ for $q=q_{max}$) gives us an estimate of the coherence length ('$\xi$'), which is of the order of reciprocal of $q_{max}$. Here our studies include both electronic mechanism and phononic mechanism.

Table 2: The calculated values of coherence length with hopping parameter of 0.2 ev and bosonic energy values of 0.3 ev for electronic and 0.05 ev for phononic mechanism respectively, under different band fillings

| Filling factor δ | Electronic mechanism Value of ξ (in unit of 'a') | Phononic mechanism Value of ξ (in unit of 'a') |
|---|---|---|
| 0.2 | 1562.5 | 2631.57 |
| 0.5 | 775.19 | |
| 0.99 | 16.67 | 111.111 |
| 1.4 | 694.44 | 3333.33 |
| 1.5 | - | 5263.16 |

Figure 2: Effect of filling factor on pairing scenario

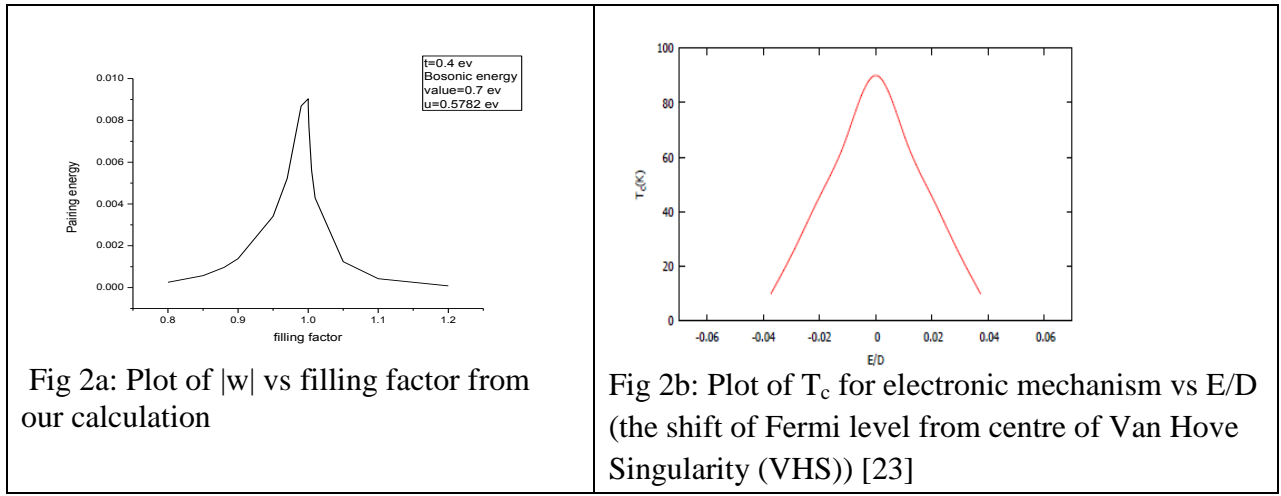

Fig 2a: Plot of |w| vs filling factor from our calculation

Fig 2b: Plot of $T_c$ for electronic mechanism vs E/D (the shift of Fermi level from centre of Van Hove Singularity (VHS)) [23]

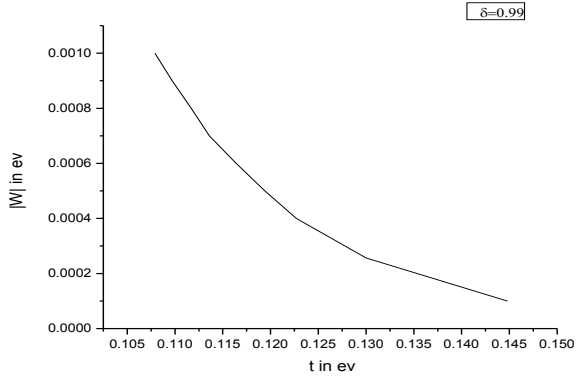

Figure 3: Variation of pairing energy with hopping amplitude at a particular filling and interaction energy (our theoretical calculations)

The relation between |W| and t, for a particular value of $\hbar\omega_{boson}$ and u at a particular $\delta$, is quite important too and our theoretical result for this is presented in Fig. 3. The graph shows that |W| increases with reduction of bandwidth. It must be kept in mind however, that below a minimum value of t the FL condition of u < 4t breaks down. So the condition sets an automatic upper bound for |W|.

Figure 4: Variation of coupling constant with bosonic energy value (both theoretical and experimental).

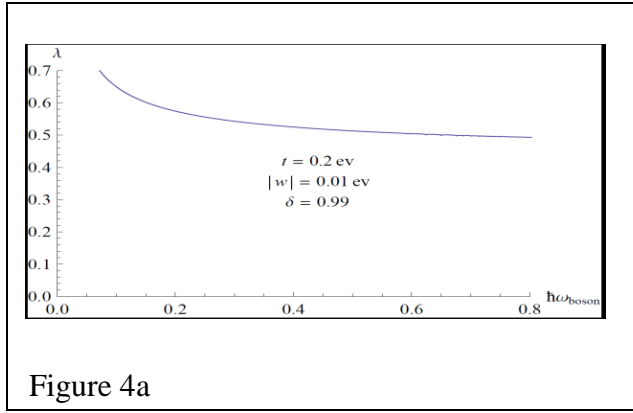
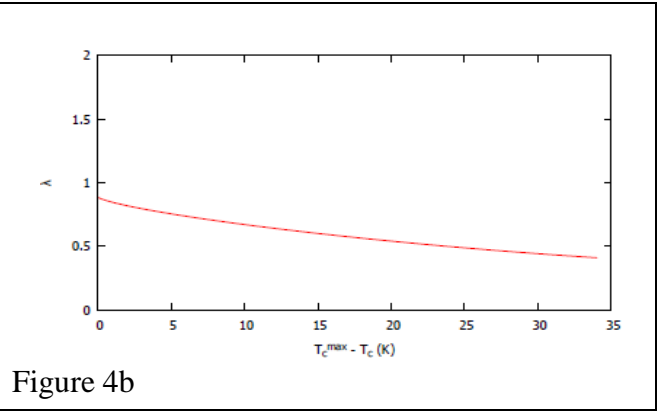

Figure 4a                                Figure 4b

A fourth (4[th]) graph is drawn to show the variation of $\lambda$ with $\hbar\omega_{boson}$ from our theoretical calculations, keeping $\delta$ the same (Fig. 4a). It should be remarked here that in Cooper's model in continuum or in BCS theory, '$\lambda$' is independent of $\hbar\omega_{boson}$. According to McMillan's equation (obtained by simplification of Eliashberg's equation) however, $\lambda$ is inversely proportional to $\hbar\omega_{boson}$ which is in qualitative agreement with the variation seen in our theoretical curve in Fig.4a [29, 30]. Again quite interestingly, the graphical nature of our above theoretical curve shows a fairly good similarity with that of an experimental curve in the overdoped phase (Fig.4b) [30]. This observation probably implies the progressive hardening of mediating boson with increase in doping in the overdoped phase.

Figure 5: Variation of reduced coherence length with filling factor (theoretical)

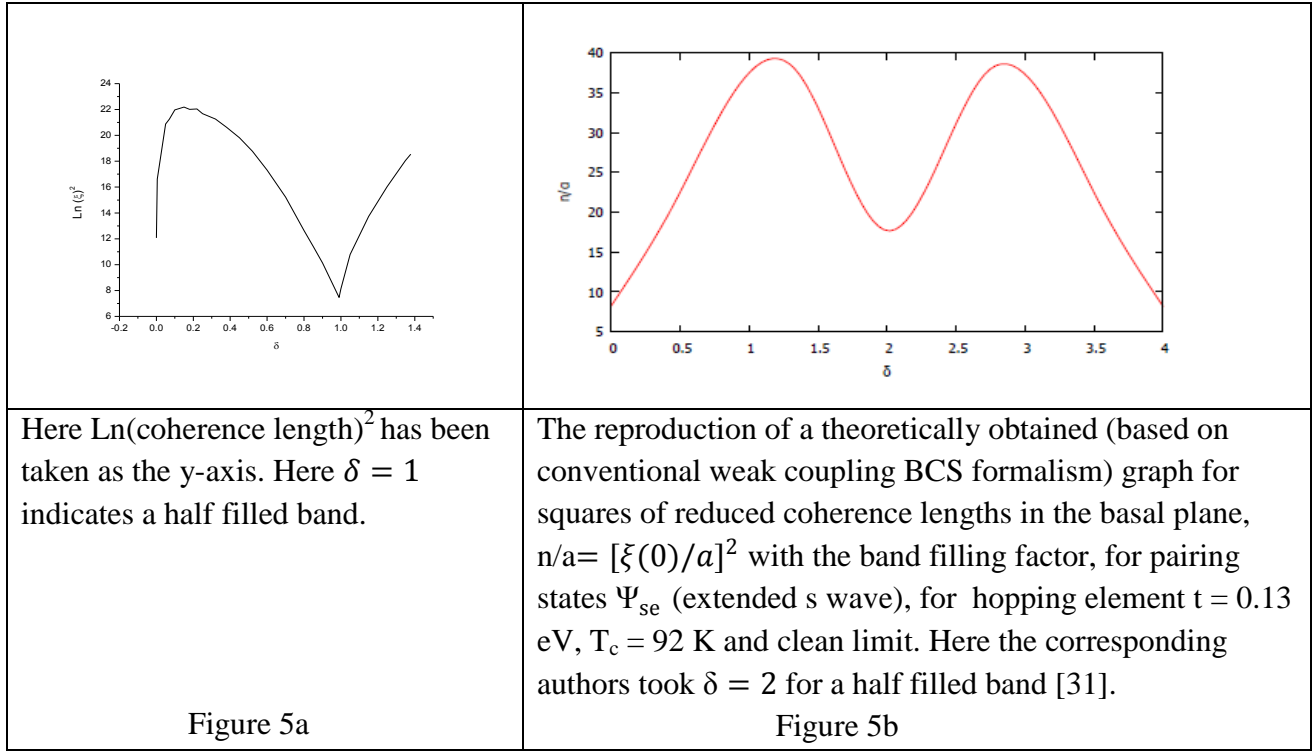

| | |
|---|---|
| Here Ln(coherence length)$^2$ has been taken as the y-axis. Here $\delta = 1$ indicates a half filled band. | The reproduction of a theoretically obtained (based on conventional weak coupling BCS formalism) graph for squares of reduced coherence lengths in the basal plane, n/a= $[\xi(0)/a]^2$ with the band filling factor, for pairing states $\Psi_{se}$ (extended s wave), for hopping element t = 0.13 eV, $T_c$ = 92 K and clean limit. Here the corresponding authors took $\delta = 2$ for a half filled band [31]. |
| Figure 5a | Figure 5b |

Finally a graph is drawn between the Log of the square of the coherence length in unit of lattice parameter ((reduced coherence length)$^2$) against band filling factor from the results of our theoretical calculations (Fig. 5a). The pairing equation (equation (1b)) becomes non tractable beyond a particular filling. So the right hand part of the graph in Fig. 5a remains unclear. Nevertheless one can see a lot of similarities between the nature of the curves in Figs 5a and 5b [31]. This study strengthens the conjecture that the coherence length is a very sensitive function of filling factor [31].

**Multiple pair (Active Fermi sea) :** The estimates for coherence length obtained in this case, is presented in the Table below.

Table 3: The calculated values of coherence length with hopping term of 0.2 ev and bosonic energy value of 0.3 ev for electronic mechanism in the overdoped region

| $\delta$ | Value of $\xi$ (in unit of 'a') |
|---|---|
| 0.5 | 666.67 |
| 0.99 | 15.65 |
| 1.4 | 692 |
| 1.5 | - |

There is no difference in the magnitudes of the coupling constant in the case of zero centre of mass momentum pairing between the situations corresponding to active Fermi sea and passive Fermi sea. This can be checked by putting q=0 in equation (1d). We present below a graph for pairing energy versus coupling constant corresponding to finite centre of mass momentum in the active Fermi sea case with a particular value of q.

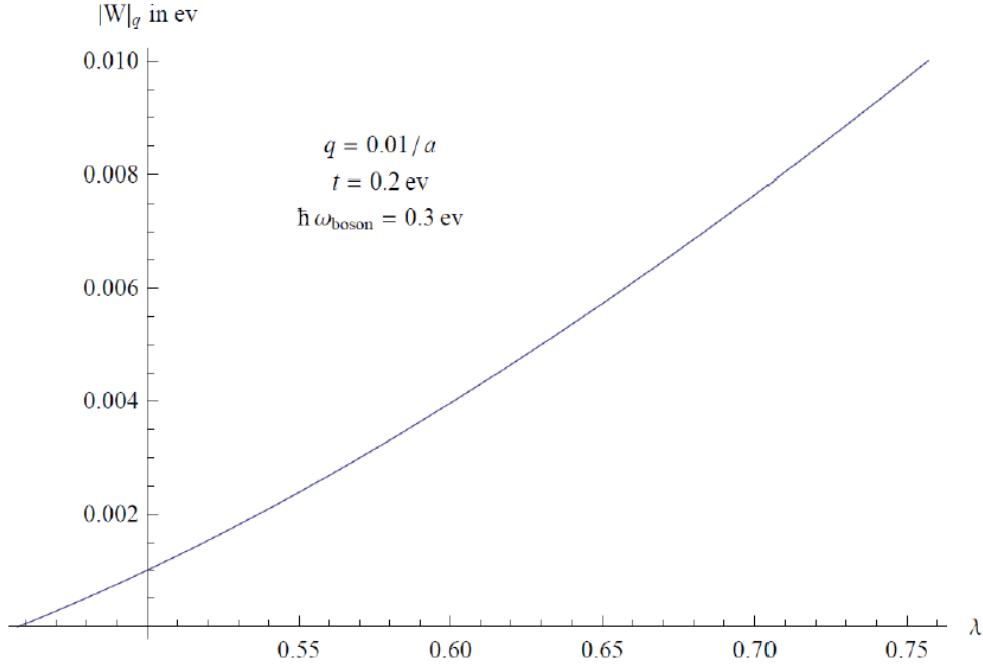

Figure 6: Pairing energy vs coupling constant for multiple pairing for a finite value of q.

Strikingly our calculations with both passive and active Fermi surfaces show the most realistic signature of pairing in moderately overdoped (intermediate band filling range) region. The coherence length is smallest and is of the order of experimental value in the vicinity of half filling. Otherwise either it is too long or the calculation becames non tractable. If we try to look into the practical significance of our result then we recall that as per the experimental phase diagram of cuprate superconductors the underdoped normal phase doesn't obey the FL theory [11]. In our calculation we get a non- tractable regime in the upper portion of the band above a particular filling for a fixed bosonic energy and hopping parameter with both passive and active Fermi sea. The extremely high doped regime in the experimental phase diagram obeys FL theory but doesn't show pairing. Similarly our calculation shows tractable results maintaining FL criterion but gives very long coherence length in low filling (extremely overdoped regime) indicating an unrealistic situation for pairing.

Generally real high- $T_c$ cuprates (both in the under and overdoped regions) have short in-plane coherence length [32, 33]. Both experimental results and our theoretical estimates (shown in

Table- 2 and Table- 3) point to the fact that the in- plane coherence lengths corresponding to excitonic mechanism are much shorter than those in the conventional phonon driven 3D superconductors. Even in 2D phononic mechanism produces a much longer coherence length (as shown in Table 2). Lowering of the magnitude of coherence length in this case needs a large enhancement in the phonon mediated attractive interaction. This violates the FL scenario and makes it inappropriate for overdoped phase. Therefore, the excitonic mechanism is the most feasible one for intra- layer pairing in the overdoped regime.

We now study the dependence of pairing energy on band filling factor. Our theoretical results are presented in Fig. 2a. It is worthwhile to point out that a numerical trick has been applied to avoid the Van Hove singularity (VHS) arising automatically in the expression for the DOS in 2D. For this, the Fermi level positions have been shifted slightly upward by $10^{-6}$ ev.

The graph in Fig. 2a shows a slow variation in $|W|$ for low $\delta$. It starts showing lattice effect at higher fillings. This curve is strikingly similar to the one for critical temperature vs filling factor obtained through Eliashberg formalism (see the graph in fig. 2b) [23, 24]. The peaks occur at around the same position (near half filling) in both the graphs. However the Hamiltonian used in the latter work includes a repulsive Coulomb term as well but the calculation involves a simpler logarithmic approximation to the general 2D DOS. Yet, the high degree of resemblance between these two graphs (Figures 2a and 2b) must be considered as quite remarkable.

The numerical results regarding multiple pairing show slight changes in the magnitude of coherence length compared to the single pair situation. It is shorter here than that obtained in the other case.

**Discussion:**

1) Although the boson mediated attractive coupling constant is believed to lie in the strong coupling regime for the superconducting cuprates in the vicinity of optimum doping concentration, there is also indication that coupling constant decreases continuously with raising doping level in the overdoped phase [30]. Our calculational results, in the overdoped region with electronic mechanism producing coupling constants in the intermediate regime, look consistent with this.

2) Analysis of the available experimental results from Angle Resolved Photoemission Spectroscopy (ARPES) and from polar angular magnetoresistance oscillations show the presence of a 3D coherent electronic behaviour in overdoped phase of some cuprate superconductors. Investigation of the oscillations shows that at certain symmetry points however, the Fermi surface exhibits properties characteristic of 2D systems [34]. This striking form of the Fermi surface topography, provides a natural explanation for a wide range of anisotropic properties both in the normal and superconducting states of this system. This result is in tune with our conjecture that in the presence of the extreme electrical anisotropy, the high-$T_c$ materials in the overdoped region can be well understood within a framework of a proper combination of 2D intra-layer pairing theory under a FL- like background and possible inter-layer pair tunneling.

Moreover, this picture in certain situations can even approximately conform to a 3D isotropic BCS treatment, as our calculations show.

3) Our calculations for Cooper's one pair problem on a 2D lattice with inclusion of band structure effects and in the background of both passive and active Fermi sea, brings out a lot of interesting physics in general agreement with other kinds of many body treatments like Eliashberg scheme and with experiments on real superconductors as well. It may also be remarked that a vertex correction can become quite important in the first principle calculation for 'u' if the bosonic energy (relevant for excitonic boson) becomes comparable to the Fermi energy [35]. This would be taken up in future.

4) Cooper pairing on lattice would in general imply momentum space pairing involving reciprocal lattice vectors as well. Restricting our calculations to one band only we however neglect these Umklapp processes.

5) Our calculations were done both with a phononic mechanism as well as an electronic one. The coupling constant regime in the two cases lie in the strong and intermediate range respectively but the magnitude of the coherence length, from the phononic process turned out to be too big for the cuprate family.From this viewpoint the excitonic mechanism seems to be the more likely one for the real overdoped cuprate superconductors. It may be noted that the presence of a d-wave symmetry in the pair wave function even in the overdoped phases of real cuprates is not ruled out. We have just examined the adequacy and feasibility of the s- wave symmetry in intra-layer pair formation.

7) To make our calculations more realistic we have considered pair formation in the background of an active Fermi sea i.e. in the presence of multiple pairs as well. Our preliminary calculational results indicate that the formation of pairs with much lower binding energy viz. of the order of $10^{-6}$ev is possible with the coupling constant in the vicinity of the intermediate regime.

8) The success of the present calculation shows that the extension of our formalism for investigation of Cooper pair formation on a 2D lattice in a generalized weakly correlated situation to that in the background of a Gutzwiller projected Fermi sea, can be very suitable for studying superconductivity in the underdoped cuprates. Our scheme may be generalized to include various kinds of mediating bosons and pairing channels as well.